\documentclass[prb,twocolumn,aps,superscriptaddress,showpacs,floatfix]{revtex4-2}
\usepackage{amsbsy,amssymb,amsmath,bm}
\usepackage{graphicx,color,epsfig,rotate}
\usepackage{xspace,units}
\usepackage{textcomp}		
\usepackage{epstopdf}
\usepackage{tabularx}

\usepackage{multirow}		
\usepackage{dcolumn}							
\newcolumntype{d}[1]{D{.}{.}{#1}}
 
\begin{document}
\title{Magnetic Forces in Paramagnetic Fluids}

\author{Tim A. Butcher}
\email{tbutcher@tcd.ie}
\altaffiliation{Present address: Swiss Light Source, Paul Scherrer Institut, 5232 Villigen PSI, Switzerland}
\affiliation{School of Physics and CRANN, Trinity College, Dublin 2, Ireland}

\author{J. M. D. Coey}
\affiliation{School of Physics and CRANN, Trinity College, Dublin 2, Ireland}

\date{\today}

\begin{abstract}

\noindent  An overview of the effect of a magnetic field gradient on fluids with linear magnetic susceptibilities is given. It is shown that two commonly encountered expressions, the magnetic field gradient force and the concentration gradient force for paramagnetic species in solution are equivalent for incompressible fluids. The magnetic field gradient and concentration gradient forces are approximations of the Kelvin force and Korteweg-Helmholtz force densities, respectively. The criterion for the appearance of magnetically induced convection is derived. Experimental work in which magnetically induced convection plays a role is reviewed.

\end{abstract}

\maketitle

\section{Introduction}
The classic hallmark of magnetic materials is their motion upon magnetisation in a magnetic field gradient. A force density known as the Kelvin force draws paramagnetic substances into increasing fields along the field gradient, whereas diamagnetic matter is repelled. A fluid constitutes a state of matter for which this is particularly apparent, due to the deformations and convective movement that magnetic forces can bring about. Examples of paramagnetic fluids range from paramagnetic gases to electrolytic solutions of paramagnetic salts, which are particularly interesting for electrochemical experiments. Paramagnetic salt solutions contain paramagnetic ions of transition metals with an unfilled d-subshell (e.g. Cr, Mn, Fe, Ni and Co) or of rare earths with unpaired 4f electrons (e.g. Gd, Tb, Dy, Ho and Er) \cite{andres_1976}. Oxygen found in its paramagnetic molecular form is an example of a paramagnetic gas \cite{selwood_book_oxygen}. Air consists of 21\% O$_2$ and is therefore also paramagnetic. The temperature dependence of their magnetic susceptibility is given by Curie's law, which also directly relates the magnetic properties to the electronic structure of the atom \cite{monzon_2014_kelvin,coey_book_2010}.

Colloidal suspensions of ferromagnetic or ferrimagnetic nanoparticles in a non-magnetic carrier liquid such as water or oil are known as ferrofluids and show a stronger response to magnetic field gradients than paramagnetic fluids based on individual non-interacting magnetic moments. Ferrimagnetic magnetite (Fe$_3$O$_4$) or maghemite ($\gamma$-Fe$_2$O$_3$) nanoparticles with sizes in the order of 10\,nm are most commonly employed in ferrofluids. Due to their small size the nanoparticles are monodomain and superparamagnetic. A ferrofluid cannot sustain a remanent magnetisation upon removal of an applied magnetic field and its magnetisation curve resembles that of a paramagnetic salt solution, albeit with a significantly higher and readily attainable saturation magnetisation. Hence, ferrofluids can be referred to as superparamagnetic liquids.

This article provides an overview of the origin of magnetically induced convection and situations in which effects of magnetic forces are relevant for experiments involving paramagnetic salt solutions and gases.

\section{Magnetic field effects on microscopic dipoles}

In order to understand the Kelvin force it is necessary to recall the forces on the magnetic moments of the atoms, ions or molecules that the fluid comprises. The microscopic force on an individual magnetic moment \textbf{m} (in A\,m$^2$) in a magnetic field gradient $\nabla \mathbf{H}$ (in A\,m$^{-2}$) is given by \cite{MIT_cont_mech_forces,zahn_ch5}: 

\begin{equation}
	\label{eq:micro_mag_force}
	\mathbf{f}_\mathrm{mag} = \mu_0 (\textbf{m} \cdot \nabla) \mathbf{H},
\end{equation}
with the vacuum permeability $\mu_0 = 4 \pi \times 10^{-7}$\,N\,A$^{-2}$. This is the expression for the force on a magnetic dipole in the magnetic charge model \cite{greene_1971, boyer_1988, griffiths_1992,yaghjian_1999}. The expression describes the situation when a single magnetic dipole is unperturbed by collision with other molecules. Thus, the measurement of this force requires a setup in which a significant free path length of the paramagnetic molecules is guaranteed. This was first achieved in the Stern-Gerlach experiment \cite{stern_1921,gerlach_1922,schmidt_2016,friedrich_2021}, which was performed with a charge neutral molecular beam \cite{estermann_1975, meerakker_2012} of paramagnetic silver atoms just over a hundred years ago. The Ag atoms entered a magnetic field gradient where they were deflected by the magnetic force before their detection. Essentially, the apparatus constituted a microscope for the momentum of the Ag atoms \cite{schmidt_2016}. A Stern-Gerlach type splitting of charged particles has never been observed, although proposals for an experimental campaign exist \cite{garraway_2002, batelaan_1997,garraway_1999,gallup_2001,garraway_2002,batelaan_2002,mcgregor_2011,henkel_2019}. An argument based on a combination of the Lorentz force and the uncertainty principle was originally made by Niels Bohr against the observability of such an effect \cite{mott_1929}. 

The mean free path in a gas is drastically reduced and common values are in the order of 70\,nm. Molecules in liquids are in close contact and react collectively due to interparticle collisions. What effect does the application of a magnetic field have on such an ensemble?

A uniform magnetic field does not exert a force on the magnetic moments, but causes them to precess. This Larmor precession is irrelevant for monatomic gases, as the collisions are independent of the orientation of the atoms. However, polyatomic gases in a homogeneous magnetic field show a decrease in thermal conductivity and viscosity of order one percent. The effect in paramagnetic polyatomic gases is known as the Senftleben effect \cite{senftleben_1930, senftleben_1933_I, senftleben_1936_II}. The extension to diamagnetic gases is called the Senftleben-Beenakker effect \cite{beenakker_1962, beenakker_1970, honeywell_1972, moraal_1972}. The modification of the transport properties relies on the averaging of the collision cross section and change of the mean free path by the precession of the magnetic moment in the field \cite{gorter_1938, zernike_1939}. The strength of the effect is proportional to $\tfrac{H}{P}$, with the pressure $P$. No equivalent of the Senftleben effect exists in liquids, where a mean free path is undefined. 

In the case of an inhomogeneous field, the microscopic force given by Eq.~\ref{eq:micro_mag_force} acts on the individual magnetic dipoles in the medium. The microscopic forces are transferred to the bulk fluid via interparticle forces that are mediated by collisions. Thus, a magnetic body force arises, which can lead to bulk motion of a magnetised fluid in a process called magnetic convection. A force density must be introduced to describe the action of the field gradient, which will be discussed in the next section.

The appearance of magnetic convection has been cast as a paradox in the past \cite{gorobets_2013, gorobets_2014, gorobets_2014_MHD, gorobets_2015_j_sol_state_echem, gorobets_2015_JAP, gorobets_2017, gorobets_2017_mol_liq, gorobets_2019}. One argument is that the discrepancy between magnetic and thermal energies is so large that it makes any magnetic modification of the movement impossible. This reasoning does not stand up to scrutiny. Velocities due to Brownian motion average out in a portion of fluid that contains an enormous number of molecules. The collisions due to the magnetic forces, on the other hand, do not. Brownian motion does not influence a macroscopic hydrodynamic flow, but force densities do. Fluid dynamics describes the situation for continuous media and the fluid is assigned average macroscopic quantities. 

In the last decade, several publications claimed to observe the enrichment of paramagnetic ions from homogeneous solutions \cite{eckert_2012, eckert_2014, yang_2014, demirors_2013, franczak_2016,kolczyk_2016, ji_2016}. This came as a surprise, because the Brownian motion of the ions enforces a uniform concentration throughout the solvent. Thermodynamically, the magnetic field hardly affects the chemical potential and the thermodynamic equilibrium remains virtually unchanged \cite{butcher_2021_jpc}. It later became clear in a set of experiments \cite{eckert_2017,rodrigues_2017,rodrigues_2019,lei_2020, lei_2021,lei_2022, fritzsche_2022} that the observed concentration changes in the field gradient could be explained by evaporation of the solvent. A completely filled and sealed cuvette does not show any inhomogeneity upon exposure to a magnetic field gradient \cite{rodrigues_2017}. Measurable concentrations of paramagnetic ions do not settle from homogeneous solutions in magnetic field gradients commonly encountered in laboratories. The density of the paramagnetic ions does not change and paramagnetic salt solutions are incompressible. However, \textit{pre-existing} concentrations of paramagnetic ions can be readily manipulated with a magnet in the time window before the system is homogenised by diffusion \cite{coey_2009,butcher_2020}. 

Unlike paramagnetic salt solutions with their individual solvated paramagnetic cations, ferrofluids consist of ferromagnetic or ferrimagnetic nanoparticles in which all the encapsulated magnetic moments are ordered by exchange interaction. The resulting total magnetic moment of such nanoparticles is in the order of 10$^4$\,$\mu_B$ (Bohr magneton: $\mu_B = 9.274 \times 10^{-24}$J\,T$^{-1}$, note that J\,T$^{-1}$ and A\,m$^2$ are equivalent units)) \cite{huke_2004}, which can be compared to the effective magnetic moment of Fe$^{3+}$ $\mu_{\mathrm{eff}}=5.9\,\mu_B$ \cite{coey_book_2010}. According to Eq.~\ref{eq:micro_mag_force}, the magnitude of the force on a magnetic nanoparticle in a magnetic field gradient is approximately 1000 times higher than that on an individual paramagnetic ion. Ferrofluids are synthesised to avoid agglomeration in a magnetic field gradient, which is ensured by the small size of the nanoparticles that is below 10\,nm in ideal ferrofluids \cite{kole_2021}. Ideal ferrofluids are incompressible. In concentrated ferrofluids, there is an additional risk of agglomeration due to magnetic dipolar and Van der Waals interactions between individual nanoparticles. The effect of these interactions can be suppressed by coating the nanoparticles with a surfactant \cite{huke_2004}. 

Without an applied magnetic field ferrofluids show no magnetisation, because the magnetic moments of the nanoparticles fluctuate. There are two reasons: firstly, rotational diffusion of the nanoparticles in the carrier liquid results in random orientations of the magnetic moments. Secondly, the thermal energy at room temperature may be high enough to cause the magnetic moment to flip randomly as is characteristic for superparamagnetic particles. The former is known as Brownian and the latter as N\'{e}el relaxation \cite{huke_2004}. When the size of the magnetic particles is increased to the microscale, the resulting liquids are known as magnetorheological fluids \cite{kumar_2021}. This material class was developed in the 1940s prior to the discovery of ferrofluids in the mid 1960s. Magnetorheological fluids experience a dramatic increase in their viscosity upon exposure to magnetic fields, but this comes to the detriment of the stability of the fluid. Brownian motion is no longer able to prevent agglomeration of the magnetic particles.

\section{Macroscopic magnetic force densities}

Magnetic field gradients exert forces on magnetically polarisable media. Derivations of expressions for the force density are outlined in many texts \cite{slepian_1950, brown_1951_I, brown_1951_II, byrne_1977, zahn_ch5, MIT_cont_mech, blums_1996, zahn_2006, lai_1981, rosensweig_2013}. In order to heuristically transform Eq.~\ref{eq:micro_mag_force} into an expression for the force density of non-interacting magnetic dipoles, scaling with their number $N$ per unit volume $V$ (number density $n = \tfrac{N}{V}$) is necessary. The magnetisation $\mathbf{M} = \tfrac{\sum_{i=1}^N  \mathbf{m}_i}{V} =  \tfrac{N  \mathbf{m}}{V} = n \mathbf{m}$ is a macroscopic variable that is defined as the number density of magnetic dipole moments multiplied by the average microscopic dipole moment in direction of the applied field. Replacing \textbf{m} with \textbf{M} in Eq.~\ref{eq:micro_mag_force}, one immediately obtains the Kelvin force expression for the magnetic force density in N\,m$^{-3}$:

\begin{equation}
	\label{eq:kevlin_force}
	\mathbf{F}_{\mathrm{K}} = \mu_0 (\mathbf{M}\cdot \nabla) \mathbf{H}.
\end{equation}

Kelvin arrived at this expression by considering the magnetic moment per unit volume and relating it to the mechanical force in a dipole model \cite{smith_1949}. The Kelvin force is often referred to as the magnetic field gradient force in magnetoelectrochemistry \cite{weston_2010,monzon_2014_kelvin}. Its intensity is proportional to the magnetic susceptibility $\chi = \tfrac{M}{H}$ of the fluid, which is around $1$ for ferrofluids and $10^{-3}$ for concentrated paramagnetic salt solutions. 

Providing the magnitude of the magnetic susceptibility is small ($\chi \ll 1$), the magnetic flux density $\mathbf{B}$ (unit: T) can be approximated as $\mathbf{B} = \mu_0 (\mathbf{H} +\mathbf{M}) = \mu_0(\mathbf{H} +\chi \mathbf{H})  \approx \mu_0 \mathbf{H}$. Consequently, the Kelvin force (Eq.~\ref{eq:kevlin_force}) becomes:

\begin{equation}
	\label{eq:grad_force}
	\mathbf{F_{\nabla \mathrm{B}}} = \frac{\chi}{\mu_0} (\mathbf{B} \cdot \nabla) \mathbf{B}.
\end{equation} 

The name of Eq.~\ref{eq:grad_force} is magnetic field gradient force and it is commonly encountered in magnetoelectrochemistry, which deals with paramagnetic salt solutions, gases or diamagnetic fluids \cite{monzon_2014_kelvin}. A version that is not scaled by $\chi$ is known as the magnetic tension force in magnetohydrodynamics (MHD) \cite{rieutord_2015_mag}. The name is aptly chosen, since it straightens out magnetic field lines.  

In the presence of current densities $\mathbf{j}$ (in A\,m$^{-2}$), a Lorentz force density $\mathbf{F_L} = \mathbf{j} \times \mathbf{B}$ must be added to the magnetic force density to account for the interaction of moving charges with the magnetic field \cite{monzon_lorentz_2014}. The Lorentz force density is of smaller magnitude than the Kelvin force when a high concentration of paramagnetic ions is present in the solution \cite{mutschke_2008,monzon_2014_kelvin}. In the following discussion, currents are ignored. 

There is a second expression for the force on a paramagnetic liquid, which incorporates the gradient of the magnetic susceptibility (or permeability):

\begin{equation}
	\label{eq:helmholtz_force}
	\mathbf{F}_{\mathrm{H}} = - \frac{\mathbf{H} \cdot \mathbf{H}}{2} \mu_0 \nabla \chi = - \frac{1}{2} H^2 \mu_0 \nabla \chi.
\end{equation}

This expression predicts a force density at interfaces where the susceptibility changes. This form of the magnetic force density was developed first by Diederik Korteweg \cite{korteweg_1880}, then by Hermann von Helmholtz \cite{helmholtz_1881} and others \cite{kirchhoff_1885, kirchhoff_1885_tube} within the framework of the thermodynamic principle of minimum energy. It is known as the Korteweg-Helmholtz force \cite{MIT_cont_mech_forces}. In magnetoelectrochemistry, the Korteweg-Helmholtz force is referred to as the concentration gradient force \cite{leventis_2005,weston_2010,svendsen_2020} with the molar magnetic susceptibility $\chi_m$ (in m$^3$\,mol$^{-1}$) and approximation $\mathbf{B} \approx \mu_0 \mathbf{H}$ for $\chi \ll 1$:

\begin{equation}
	\label{eq:conc_grad_force}
	\mathbf{F}_{c} = - \frac{1}{2} H^2 \mu_0 \chi_{m} \nabla c \approx - \frac{B^2}{2\mu_0} \chi_{m} \nabla c.
\end{equation}

On the face of it, the situation seems bewildering. The Kelvin force (Eq.~\ref{eq:kevlin_force}) and the Korteweg-Helmholtz force (Eq.~\ref{eq:helmholtz_force}) predict force densities in inhomogeneous magnetic fields or at interfaces at which the susceptibility changes, respectively. There has been controversy over which of the expressions is correct ever since they were first derived. The argument has not abated to the present day \cite{gingras_1980, coey_2007,grinfield_2015, svendsen_2020}. Criticism was first levelled at the Korteweg-Helmholtz force by followers of Kelvin \cite{larmor_1897, livens_1916, smith_1949}. 

The apparent paradox can be resolved by re-expressing the Kelvin force density (Eq.~\ref{eq:kevlin_force}) with the help of vector identities \cite{MIT_cont_mech_forces}. The first step is to expand the expression with $(\mathbf{H} \cdot \nabla) \mathbf{H} = (\nabla \times \mathbf{H}) \times \mathbf{H} + \tfrac{1}{2} \nabla (\mathbf{H} \cdot \mathbf{H})$ after writing $\mathbf{M} = \chi \mathbf{H}$:


\begin{equation}
	\mathbf{F}_\mathrm{K} = \mu_0 \, \chi [(\underbrace{\nabla \times \mathbf{H})}_{=0} \times \mathbf{H} + \tfrac{1}{2} \nabla(\mathbf{H}\cdot \mathbf{H})].
\end{equation}

In the absence of currents $\nabla \times \mathbf{H} = 0$ and the first term disappears. The remaining dot product of $\mathbf{H}$ is already redolent of the Korteweg-Helmholtz force (Eq.~\ref{eq:helmholtz_force}) and can be further transformed by employing $\nabla (\chi \mathbf{H}^2) = \chi \nabla \mathbf{H}^2 + \mathbf{H}^2 \nabla \chi$:

\begin{equation}
	\mathbf{F}_\mathrm{K} = \nabla\underbrace{\left( \frac{\mu_0}{2} \chi \mathbf{H} \cdot \mathbf{H}\right)}_{P_\mathrm{mag}}  - \frac{\mu_0}{2} \mathbf{H} \cdot \mathbf{H} \, \nabla\chi. 
	\label{eq:kelvin_pressure}
\end{equation}

The term in the brackets on the left is a magnetic readjustment of the internal pressure in the fluid $P_\mathrm{mag}$. This is not present in the Korteweg-Helmholtz version of the force density \cite{rosensweig_2013,mazur_1956} and instead appears within the definition of the internal pressure itself. The gradient of the magnetically induced pressure is the difference between the two expressions for the magnetic force density on a paramagnetic fluid:

\begin{equation}
	\mathbf{F}_\mathrm{K} = \nabla P_\mathrm{mag} + \mathbf{F}_\mathrm{H}.
\end{equation}

Formulation in terms of the magnetic field gradient force and the concentration gradient force is also possible: 

\begin{equation}
	\mathbf{F_{\nabla \mathrm{B}}} = \nabla P_\mathrm{mag} + \mathbf{F}_\mathrm{c}.
\end{equation}

What are the implications of this algebra for the dynamics of a paramagnetic fluid element that is exposed to a magnetic field gradient? Gradients of the pressure are unimportant for the prediction of deformation of a portion of incompressible fluid. Only rotational forces can cause fluid movement which deforms the paramagnetic solution in a closed system in which the fluid is bounded by solid walls \cite{shercliff_1979}. The gradients of internal pressure are by definition irrotational and inconsequential for deformations that leave the volume unchanged. It follows that the Kelvin and Korteweg-Helmholtz expressions for the magnetic force density on a paramagnetic fluid predict the same motion \cite{MIT_cont_mech_forces,mazur_1956,mazur_1956,lahoz_1980,brevik_1982}, despite the fact that they lead to different distributions of the force density. Deformations of incompressible fluids, such as paramagnetic salt solutions, are equally well described by \textit{either} expression. For compressible paramagnetic fluids, the magnetic readjustment of the pressure is relevant and can cause magnetostriction \cite{rosensweig_2013,zahn_2006}. In this case, the magnetic contribution to the pressure and density must be accounted for as was shown in dielectric liquids \cite{hakim_1962,lahoz_1980,brevik_1982}.

The futility of the argument about the correct force density was first pointed out in the 1950s \cite{brown_1951_II, mazur_1956, mazur_book_kelvin}. The reasoning was later presented again \cite{byrne_1977, lahoz_1980,brevik_1982,park_2020}. An interpretation of the effects of the magnetic force densities must be sought in fluid dynamics, which is the aim of the next section.

\section{Magnetically induced convection}

The Navier-Stokes equation is fundamental for fluid dynamics. It relates the acceleration of the fluid, given by the material derivative of the fluid velocity ($\tfrac{\mathrm{D}\mathbf{u}}{\mathrm{D}t} = \tfrac{\partial \mathbf{u}}{\partial t} + \mathbf{u} \cdot \nabla \mathbf{u}$), to force fields and pressure gradients. The Navier-Stokes equation for a paramagnetic fluid can be written with the Kelvin force:

\begin{equation}
	\label{eq:nav_stokes}
	\rho \frac{\mathrm{D}\mathbf{u}}{\mathrm{D}t} = - \nabla P + \eta \nabla^2 \mathbf{u} + \rho \mathbf{g} + \mu_0 (\mathbf{M} \cdot \nabla) \mathbf{H}, 
\end{equation}

\noindent with the pressure $P$, dynamic viscosity $\eta$ (in N\,s\,m$^{-2}$), the density $\rho$ and the gravitational acceleration $\mathbf{g}$. The two last terms on the right are the gravitational and Kelvin force densities. Generally, the pressure is unknown and it is impossible to make a direct prediction of fluid motion from Eq.~\ref{eq:nav_stokes} without the incompressibility constraint $\nabla \cdot \mathbf{u} = 0$. The issue can be sidestepped by applying the curl operator ($\nabla \times$) to the Navier-Stokes equation, disposing of $\nabla P$ \cite{shercliff_1979}. This brings into play the vorticity of the flow. Any body force with a non-zero curl is able to deform the fluid and causes it to rotate \cite{tritton_2012_vorticity}. If the flow is irrotational, a velocity potential $\mathbf{u} = \nabla \phi$ can be introduced and the situation is known as potential flow. Any irrotational (potential) force is balanced by the pressure field in the liquid. These pressure changes are equilibrated by solid walls in a closed cell and the fluid remains static in the absence of free surfaces. The pressure serves as an assurance that only deformations in which the volume is maintained are allowed in an incompressible fluid. No appreciable changes in concentration are to be expected from commonly encountered pressure differences. Only rotational body forces can create internal flows.

In order for paramagnetic fluids to be moved by magnetic convection, the Kelvin force must be rotational. So what is the condition for the non-potentiality of the Kelvin force and the appearance of magnetically induced convection? The prerequisites can be obtained by deriving the condition for mechanical equilibrium from Eq.~\ref{eq:nav_stokes}. Any magnetised fluid undergoes convective movement with the aim to establish this mechanical equilibrium.  Under zero flow ($\mathbf{u}=0$) the application of $\nabla \times$ to Eq.~\ref{eq:nav_stokes} yields:

\begin{equation}
	\label{eq:curl_nav}
	\nabla \times [\rho \mathbf{g} + \mu_0 (\mathbf{M \cdot \nabla}) \mathbf{H}] = 0.
\end{equation}

With the vector identity $\nabla \times (\psi \mathbf{A}) = \psi \nabla \times \mathbf{A} + \nabla \psi \times \mathbf{A}$ and using the collinearity of $\mathbf{M}$ and $\mathbf{H}$ this becomes:

\begin{equation}
	\label{eq:kelv_deriv}
	\nabla \rho \times \mathbf{g} + \mu_0 \nabla \times \left(\frac{M}{H} \mathbf{(H \cdot \nabla)} \mathbf{H}\right) = 0.
\end{equation}

The vector identity $(\mathbf{H} \cdot \nabla) \mathbf{H} = \underbrace{(\nabla \times \mathbf{H})}_{=0} \times \mathbf{H} + \tfrac{1}{2} \nabla H^2$ and the previously used vector identity for the curl lead to the criterion of hydrostatic equilibrium for magnetic fluids:

\begin{equation}
	\label{eq:mech_equil}
	\nabla \rho \times \mathbf{g} + \mu_0 \nabla M \times \nabla H = 0.
\end{equation}

Any departure from this condition leads to magnetic convection \cite{blums_1987, blums_1996}. The magnetic field gradient modifies density difference driven convection and pulls the paramagnetic fluid into the external field gradient. Mathematically, the curl of the Kelvin force causes non-potentiality of gravity by introducing density gradients orthogonal to the direction of gravity. This proceeds until Eq.~\ref{eq:mech_equil} is satisfied. 

The first term ensures stability in the gravitational field. Gravity induces convection if there is a density gradient that is non-parallel to the direction of the gravitational acceleration. For this reason, a column of water does not simply undergo an internal downward flow due to gravity. 

The second term is due to the interaction of the magnetic field gradient with the paramagnetic fluid. Gradients in both the magnetisation $\nabla M$ and the external magnetic field $\nabla H$ must be present to drive convection. Simply applying a magnetic field gradient to homogeneous paramagnetic fluids has no effect, because the magnetic pressure is cancelled by the walls. A gradient in the concentration of the paramagnetic species must be present in the system, which is accompanied by a gradient in density.

Furthermore, the gradients of $\nabla M$ and $\nabla H$ must be non-parallel for a flow to appear. This requirement is straightforward to achieve since Gauss's law for magnetism dictates a divergence-free magnetic flux density: 

\begin{equation}
	\nabla \cdot \mathbf{B} = 0.
\end{equation} 

Hence, it is impossible to generate a magnetic field gradient exclusively in a single dimension.

It is equally valid to derive the criterion for mechanical equilibrium in magnetic fluids with the Korteweg-Helmholtz force (Eq.~\ref{eq:helmholtz_force}) instead of the Kelvin force in Eq.~\ref{eq:nav_stokes}. This also results in the condition given by Eq.~\ref{eq:mech_equil}, as the Kelvin force and the Korteweg-Helmholtz force differ by a gradient of the magnetic pressure (see Eq.~\ref{eq:kelvin_pressure}). The gradient of the pressure is irrotational and drops out.

In magnetoelectrochemistry it is useful to analyse the curl of the magnetic field gradient force (Eq.~\ref{eq:grad_force}) to determine whether convection takes place in a paramagnetic salt solution. An expression with the molar magnetic susceptibility and the concentration can be obtained:

\begin{equation}
	\label{eq:curl_mag_force}
	\nabla \times \mathbf{F}_{\nabla \mathrm{B}} = \frac{1}{2\mu_0} \chi_m \nabla c \times \nabla B^2,
\end{equation}

\noindent which is simply a reformulation of the magnetic part of Eq.~\ref{eq:mech_equil}.

%

The following section will now provide a review of studies in which magnetically induced convection was observed. The focus will lie on experiments with paramagnetic salt solutions and gases. For discussions of the applications of ferrofluids, the reader is referred to existing review articles \cite{kole_2021,moskowitz_1990,torres_2014,zhang_2019_adv_mat,oehlsen_2022}.

%
%
%
%
%
%
%
%
%


\section{Overview of Experimental Observations of Magnetically Induced Convection}

Magnetic convection in a paramagnetic fluid occurs as long as the condition given by Eq.~\ref{eq:mech_equil} is not satisfied. Necessary requirements are a magnetic field gradient and a non-collinear gradient of the magnetisation. The gradient in the magnetisation is due to an inhomogeneity in the concentration of paramagnetic species. The focus of the first part of this section lies on systems in which this inhomogeneity is caused by the input of thermal energy. The second part deals with the introduction of electrical energy to electrolytic solutions under magnetic fields in magnetoelectrochemistry. 

The classical way to provoke convection is to create density differences by heating the fluid. Temperature gradients in the magnetic fluid can modify both the magnetisation and the density of paramagnetic fluids. This is the case in magnetothermal convection \cite{carruthers_1968, ueno_1987, wakayama_1991, uetake_1999,maki_2005, braithwaite_1991,bednarz_2005,bednarz_2009_I,bednarz_2009, mogi_2003, qi_2004,clark_1977, honeywell_1978, bednarz_2003, song_mag_conv_2019, huang_1998}. 

Magnetothermal convection was first investigated in gaseous paramagnetic oxygen \cite{selwood_book_oxygen,carruthers_1968, clark_1977, honeywell_1978, ueno_1987, wakayama_1991, uetake_1999}. It even found application in the measurement of O$_2$ levels in high altitude flights in the 1940s \cite{klauer_1941,dyer_1947, carruthers_1968}. When O$_2$ is heated, the magnetic susceptibility ($\chi = 0.145\times 10^{-6}$ at 20\textdegree{}C \cite{wills_1924}) decreases according to $T^{-2}$. This is due to the combined effect of the Curie law ($\chi = \tfrac{C}{T}$) and the expansion of O$_2$ ($\rho \propto T^{-1}$). Thus, the warm gas is pushed out of a field gradient and an inhomogeneous magnetic field increases the thermal conductivity by this convective motion. 

The magnetic convection of oxygen can also lead to the phenomenon of \textquotedblleft magnetic wind\textquotedblright \cite{selwood_book_oxygen,ueno_1987,wakayama_1991, uetake_1999}. A related effect may be the reported enhancement of the evaporation rate of water in magnetic field gradients \cite{nakagawa_1999, guo_2012}, which has been ascribed to the magnetic convection of diamagnetic water vapour in the paramagnetic air above the water surface \cite{nakagawa_1999}.

Later experimental studies concentrated on magnetically modified heat transfer in paramagnetic liquids, with gadolinium nitrate solution as a model system in the magnetic field of superconducting magnets \cite{braithwaite_1991, maki_2005, bednarz_2005, bednarz_2009, bednarz_2009_I}. The susceptibility of an aqueous one-molar solution of Gd$^{3+}$ ions is $\chi = 321 \times 10^{-6}$ \cite{coey_2009}.  Small permanent magnets are also sufficient to drive magnetothermal convection of concentrated paramagnetic salt solutions \cite{rodrigues_2019}.

It is possible to magnetically force convection in diamagnetic water ($\chi = -9 \times 10^{-6}$) \cite{ueno_1994,mogi_2003}. Heat transfer in regular diamagnetic water can be modified by intense magnetic field gradients, which can be provided by specialised hybrid superconducting magnets \cite{mogi_2003}. A number of numerical studies that describe the phenomena underpinning the observed heat transfer in non-conducting fluids exist \cite{huang_1998, bednarz_2003, song_mag_conv_2019}.

Attempts to use magnetothermal convection to separate rare earth ions were made by Ida and Walter Noddack in the 1950s \cite{noddack_1952, noddack_1955,noddack_1958}. They applied a magnetic field gradient to a Clusius-Dickel separation column in which the liquid is heated by Joule heat generated from a current-carrying wire \cite{clusius_1938,clusius_1939,clusius_1939_II,clusius_1939_trennrohr,grodzka_1977,muller_1988,brandon_2021} and reported an amplification of the rare earth separation by thermodiffusion \cite{noddack_1955,noddack_1958}.

Magnetic field gradients of permanent magnets can also act on paramagnetic salt solutions in which the solvent is allowed to evaporate and the top layer of the fluid gains a higher concentration in a distillation process. Several experimental works showed that placing a magnet on top of an insufficiently sealed cuvette containing paramagnetic salt solutions levitates the layer of higher concentration above the bulk fluid \cite{eckert_2017,rodrigues_2017,rodrigues_2019,lei_2020, lei_2021,lei_2022, fritzsche_2022}. Without a magnet close to the liquid surface, this layer sinks in a Rayleigh-Taylor instability and mixes with the bulk solution.   

The crystallisation of paramagnetic salts from concentrated solutions can be facilitated by confinement in a field gradient. This was first demonstrated for Mohr's salt \cite{schieber_1967} and later for nickel sulfate \cite{poodt_2005, poodt_2006}.
Magnetic field gradients have also been reported to enhance the separation of rare earths by crystallisation \cite{higgins_2020,fan_2022,kumar_2022}.

Inhibition of regular convection by trapping magnetic fluids in magnetic field gradients may also become relevant in the field of microfluidics \cite{pamme_2006,nguyen_2012micro}, because the magnetic field gradient can stabilise liquid within liquid tubes \cite{coey_2009, dunne_2020}. These dispense with the friction encountered at solid walls. The concept of magnetic control of liquid-in-liquid flow was originally developed with paramagnetic salt solutions \cite{coey_2009} and later extended to ferrofluids to maximise the confinement in the magnetic field gradient \cite{dunne_2020}. It is also possible to halt double diffusive convection in multicomponent systems containing a paramagnetic species \cite{butcher_2020}.

In electrochemistry, an input of electrical energy to electrolytic solutions via electrodes drives chemical reactions and mass transport of ions. The transport of the electrolyte to the electrode surface is strongly influenced by bulk movement of the fluid. This is where magnetic body forces can have noticeable effects, because diffusion limited concentration gradients appear close to the electrodes. 

The first explorations of magnetic field effects on electrochemical cells were carried out with homogeneous fields and focused on the Lorentz force \cite{heilbrun_1904, mohanta_1972, mohanta_1974, aogaki_1975_1,aogaki_1975_2,aogaki_1976}. This work laid the foundation for what is now known as the magnetohydrodynamic (MHD) effect, namely the stirring of the electrolyte solution by the Lorentz force close to the electrode \cite{monzon_lorentz_2014}. A magnetic field parallel to the electrode surface is orthogonal to the current and the electrolytic solution rotates. A direct consequence of this is that the limiting current is increased \cite{mohanta_1972,mohanta_1974,aaboubi_1990, hinds_1998, hinds_2021_lorentz, O_Reilly_2001, chopart_2002, fricoteaux_2003, weier_2005,weier_2007,muehlenhoff_2012}, as the convective flow helps to replenish the concentration at the electrode \cite{mutschke_2008}.

When strong magnetic field gradients $\nabla B \approx 100$\,T\,m$^{-1}$ and paramagnetic species are present in the solution, the Lorentz force is insignificant and the Kelvin force becomes dominant as soon as a gradient of paramagnetic ions is established. The classical and much studied example in which paramagnetic ions can be influenced by the Kelvin force is the electrodeposition of copper \cite{ismail_1979, hinds_2001, tschulik_2009, tschulik_2010_ECS, tschulik_2010, mutschke_2010,  dunne_2011,dunne_2012_mhd, dunne_2012_jap, dunne_2012, murdoch_2018, huang_2021, marinaro_2021, huang_2022, huang_2022_magnetochem}. A one-molar aqueous solution of Cu$^{2+}$ has $\chi = 7 \times 10^{-6}$ \cite{coey_2009}. One of the first reports of structuring copper electrodeposits from CuSO$_4$ solutions with inhomogeneous magnetic fields was reported in 1979 \cite{ismail_1979}. Since then there has been an increase in the obtainable field gradient, with experiments relying on small permanent magnets or the stray field of a magnetised wire to deliver high magnetic field gradients. The electrodeposition in a magnetic field gradient is accompanied by the formation of both gradients in the density and the magnetic susceptibility. This means that the interplay of the gravitational and the Kelvin force densities re-establishes mechanical equilibrium according to Eq.~\ref{eq:mech_equil} \cite{mutschke_2010}. 

The observation of this structuring is not limited to paramagnetic Cu$^{2+}$ solutions. Any paramagnetic ion species that undergoes reaction at an electrode causes a concentration gradient with respect to the bulk solution. An example of an organic compound that can be captured in a magnetic field gradient after electrochemical conversion is nitrobenzene \cite{ragsdale_1998, grant_1999, pullins_2001, grant_2002, leventis_2005}. It is also possible to draw paramagnetic oxygen close to an electrode backed by a magnetic field gradient and enhance the current in oxygen reduction reactions \cite{wakayama_2001, kishioka_2001, okada_2003, chaure_2007, chaure_2009}.

An important class of fluid dynamical phenomenon is flow separation where the fluid velocity changes direction with respect to that of the main fluid. This frequently happens in the vicinity of solid obstacles in the way of the flow, but it can also be triggered by magnetic field gradients \cite{mutschke_2012_comment}. A necessary requirement for separation to occur is that the vorticity of the flow changes sign \cite{tritton_2012_separation}. In the case of a solid wall, the fluid passing in immediate proximity to the obstacle is spun around and forms a counterflow. In the magnetic case, changes in the vorticity can be precipitated by variations in the curl of the Kelvin force (Eq.~\ref{eq:mech_equil}).

A phenomenon due to this magnetically induced separation of flow is the inverse structuring of electrodeposits in the intense magnetic field gradients of permanent magnets or magnetised Fe wires \cite{tschulik_2011,tschulik_2012, dunne_2011,dunne_2012_mhd,dunne_2012_jap,dunne_2012}. In these experiments, the electrochemical cell was filled with a solution containing non-magnetic, but electroplatable ion species (such as Bi$^{3+}$ or Zn$^{2+}$) and a paramagnetic ion species that was non-electroplatable under the employed experimental conditions (such as Mn$^{2+}$ or Dy$^{3+}$). When an electrodeposition from this mixture was carried out above an array of small Nd-Fe-B permanent magnets, the non-magnetic species exhibited much thicker deposits in the area of low magnetic ($\mathbf{B} \cdot \nabla) \mathbf{B}$. The presence of the paramagnetic species caused the formation of zones into which the transport of electroactive ions from the bulk solution was inhibited. The explanation of this effect was provided by analysing the curl of the magnetic field gradient force (Eq.~\ref{eq:curl_mag_force}) \cite{tschulik_2012, mutschke_2012_comment}. 
 
Recently, the improvement of water splitting by the application of a magnetic field has attracted interest \cite{garces_2019, karatza_2021,ren_2021, li_2021,xiong_2021}. Numerous studies report an increase of the measured current when a magnetic field is applied to the electrodes. A popular explanation for this observation is that the triplet state (\textuparrow\textuparrow) of molecular oxygen is favoured and therefore the magnetic field  promotes the oxygen evolution reaction. As of yet, there is no clear experimental evidence to support this interpretation. The magnetic fields in these experiments were usually inhomogeneous. For instance in \cite{garces_2019}, an increase in current was observed when a Nd-Fe-B magnet was placed behind the magnetic electrodes. The authors showed that the effect persisted when the solution was stirred and concluded that it was not a mass transport effect. Past experiments showed that the magnetic field gradient force can act in the inverse oxygen reduction reaction even when the electrode is rotated \cite{okada_2003, chaure_2009}. It is possible that the magnetic field facilitates the removal of gas bubbles from the electrode due to the Lorentz force \cite{wang_2020,li_2021_bubbles,koza_2011,liu_2019_bubbles} and this is what is increasing the current.

\section{Conclusion}

The body force caused by magnetic field gradients can induce convection in inhomogeneous systems of paramagnetic fluids. This leads to a situation in which the Kelvin force becomes rotational, which modifies mechanical equilibrium. The two common expressions for the force density in magnetoelectrochemistry are the magnetic field gradient force and the concentration gradient force. These are approximations of the Kelvin and Korteweg-Helmholtz force, respectively. Both of these accurately describe the effect of magnetic field gradients on incompressible fluids with linear magnetic susceptiblities. Miniaturisation of the magnetic field configurations gives rise to much higher gradients, which will dramatically alter magnetically induced convection in the future. Time will tell what effect magnetic forces can have on separation processes, heat transfer, transport in porous materials or electrochemical reactions.

\section{Acknowledgements}

Support from the European Commission under contract No 766007 for the MAMI Marie Curie International Training Network is acknowledged.

\bibliography{parafluid_rev}

\end{document}